\documentclass[12pt]{article}
\usepackage[british]{babel}
\usepackage{graphicx}
\usepackage{txfonts}

\newcommand\apj{ApJ}
\newcommand\aap{A\&A}
\newcommand\araa{ARA\&A}

\newcommand\nat{Nature}
\newcommand\apjl{ApJL}

\newcommand\ssr{Space~Sci.~Rev.}

\newcommand\pasp{PASP}

\newcommand\pubnumber{Article 32 in eConf C1304143}
\newcommand\pubdate{\today}

\def\napoli{
$^1$MPE;
$^2$TLS;
$^3$Mullard Space Science Laboratory;
$^4$Liverpool John Moores University;
$^5$University of Nevada;
$^6$University of Maryland Baltimore County;
$^7$CRESST;
$^8$Czech Technical University in Prague;
$^9$NASA;
$^{10}$Pennsylvania State University;
$^{11}$DARK;
$^{12}$ESO
}

\def\Title#1{\begin{center} {\Large #1 } \end{center}}
\def\Author#1{\begin{center}{ \sc #1} \end{center}}
\def\Address#1{\begin{center}{ \it #1} \end{center}}

\newcommand\pubblock{\rightline{\begin{tabular}{l} \pubnumber\\
         \pubdate  \end{tabular}}}
\newenvironment{Abstract}{\begin{quotation}  }{\end{quotation}}
\newenvironment{Presented}{\begin{quotation} \begin{center} 
             PRESENTED AT\end{center}\bigskip 
      \begin{center}\begin{large}}{\end{large}\end{center} \end{quotation}}

\def\beq{\begin{equation}}
\def\eeq#1{\label{#1}\end{equation}}
\def\eeqn{\end{equation}}

\def\beqa{\begin{eqnarray}}
\def\eeqa#1{\label{#1}\end{eqnarray}}
\def\eeqan{\end{eqnarray}}

\let\bar=\overbar

\def\Dslash{\not{\hbox{\kern-4pt $D$}}}
\def\dslash{\not{\hbox{\kern-2pt $\del$}}}

\def\msb{{\bar{\ssstyle M \kern -1pt S}}}

\begin{document}
\begin{titlepage}
\pubblock

\vfill
\Title{Simultaneous optical/gamma-ray observations of GRB 121217's prompt emission}
\vfill
\Author{J. Elliott$^1$, H. -F. Yu$^1$, S. Schmidl$^2$, J. Greiner$^1$, D. Gruber$^1$, S. Oates$^3$, S. Kobayashi$^4$, B. Zhang$^5$, J. R. Cummings$^{6,7}$, R. Filgas$^8$, N. Gehrels$^9$, D. Grupe$^{10}$, D. A. Kann$^1$, S. Klose$^2$, T. Kr\"uhler$^{11,12}$, A. Nicuesa Guelbenzu$^1$, A. Rau$^1$, A. Rossi$^2$, M. Siegel$^{10}$, P. Schady$^1$, V. Sudilovsky$^1$, M. Tanga$^1$, K. Varela$^1$}
\Address{\napoli}
\vfill
\vspace{-0.19cm}
\begin{Abstract}
Since the advent of the Swift satellite it has been possible to obtain precise localisations of GRB positions of sub-arcsec accuracy within seconds, facilitating ground-based robotic telescopes to automatically slew to the target within seconds. This has yielded a plethora of observational data for the afterglow phase of the GRB, but the quantity of data ($<$2KeV) covering the initial prompt emission still remains small. Only in a handful of cases has it been possible obtain simultaneous coverage of the prompt emission in a multi-wavelength regime (gamma-ray to optical), as a result of: observing the field by chance prior to the GRB (e.g. 080319B/naked-eye burst), long-prompt emission (e.g., 080928, 110205A) or triggered on a pre-cursor (e.g., 041219A, 050820A, 061121). This small selection of bursts have shown both correlated and uncorrelated gamma-ray and optical light curve behaviour, and the multi-wavelength emission mechanism remains far from resolved (i.e. single population synchrotron self-Component, electron distributions, additional neutron components or residual collisions). Such multi-wavelength observations during the GRB prompt phase are pivotal in providing further insight on the poorly understood prompt emission mechanism. We add to this small sample the Swift burst 121217A that had two distinct periods of prompt emission separated by ∼700 s, observed by Swift/BAT, Swift/XRT and Fermi/GBM. As a result of the time delay of the second emission, it enabled optical imaging (from 3 to 7 bands) to be taken with the GROND instrument to a resolution as fine as 10s. This multi-wavelength data will hopefully allow us to shed more light on the current picture of prompt emission physics.
\end{Abstract}
\vfill
\begin{Presented}
Huntsville Gamma-Ray Burst Symposium\\
Tennessee, USA,  April 14--18, 2013
\end{Presented}
\vfill
\end{titlepage}
\def\thefootnote{\fnsymbol{footnote}}
\setcounter{footnote}{0}

\section{Introduction}
Ever since $\gamma$-ray bursts were first detected in the 1960s~\cite{Klebesadel73a}, several satellites have been launched to expand our understanding of the underlying mechanism that caused them. One of the most notable, {\it Swift}, has detected over a thousand long duration GRBs and acquired many prompt emission light curves and spectra. Even though this huge data set has answered many questions about the GRB phenomena, the underlying problem of the prompt emission mechanism remains elusive~\cite{Zhang12a}.

The currently favoured model of the prompt emission is the internal shock scenario~\cite{Rees94a}, whereby shells of different Lorentz factors cross one another, causing relativistic shocks. The Fermi accelerated electrons then cool via synchrotron radiation~\cite{Meszaros02a, Zhang11a}. The powerlaw model predicted from the internal shock model, in the majority of cases~(\cite{Zhang11a}), does not fit the data as well as other functions, especially that of the Band model~\cite{Band93a}.

Given the current descrepencies of the favoured model, it is of utmost importance to obtain multi-wavelength studies of the prompt episodes of GRBs to test their validity. Unfortunately, given the delay between the triggering of $\gamma$-ray telescopes and the slewing of optical instruments, it is not an easy task. So far, there exist tens of fortuous cases in which both the $\gamma$-ray emission and optical emission have been detected during the prompt period. These boil down to three possible scenarios: (i) a wide-field camera is observing the same field position as a satellite and so catches the optical emission simultaneously~(e.g., 080319B;~\cite{Racusin08a,Bloom09a,Beskin10a}), (ii) the prompt period is long enough that optical instruments slew in time to observe the prompt period(e.g., 990123, 080928, 110205A, 091024;~\cite{Cucchiara11a, Gruber11b, Gendre12a, Zheng12a}), (iii) there is a precursor to the main event so that optical instruments can slew in time (e.g., 041219A, 050820A, 061121;~\cite{Blake05a, Genet07a, PageK07a}). Given the difficulty of obtaining these optical wavelength measurements, especially in multiple filters, and the low number of test cases, more bursts of this type are required.

\section{Observations}

\begin{figure}[htb]
  \centering
  \includegraphics[width=11cm]{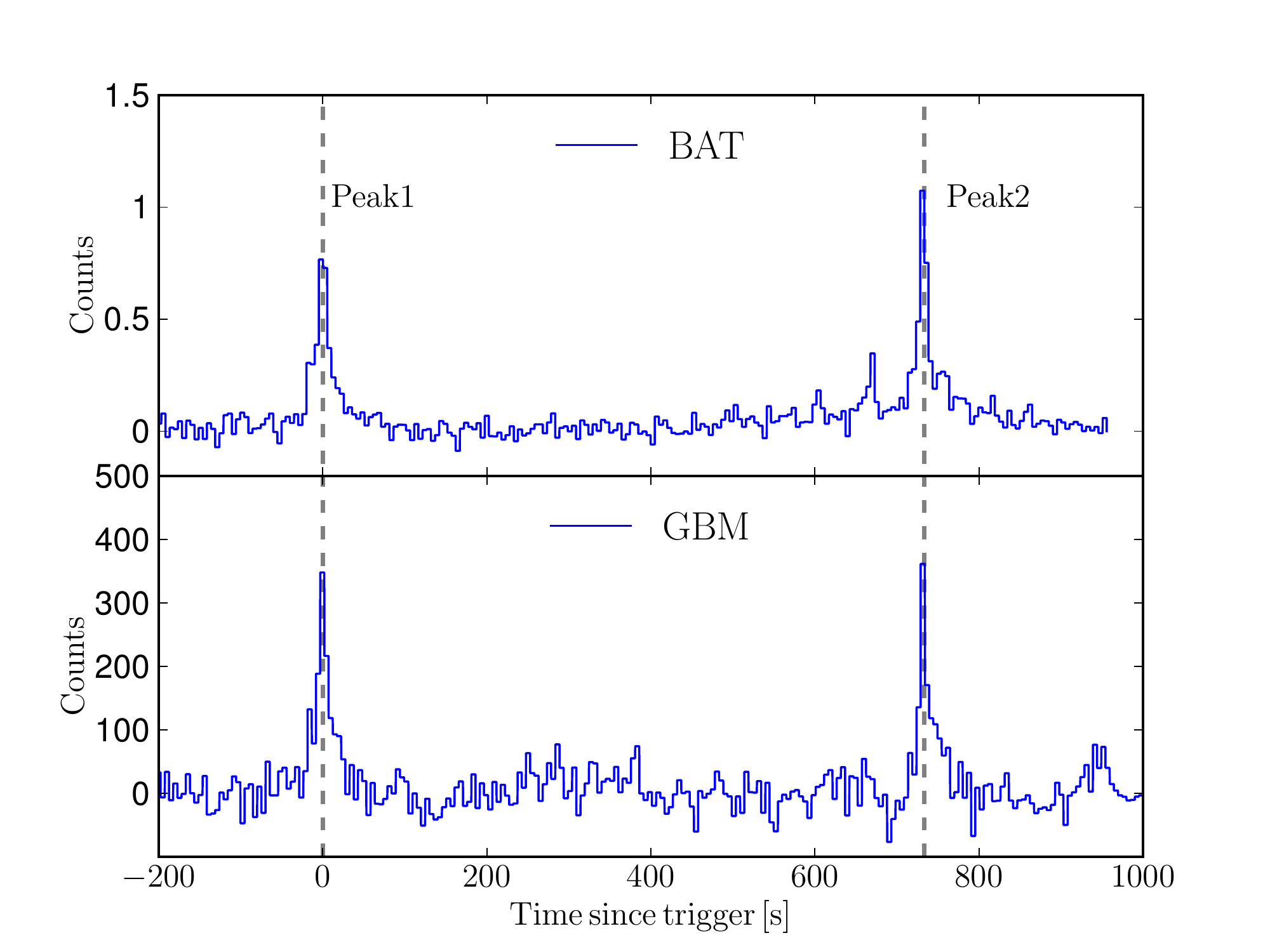}
  \caption{The $\gamma$-ray light curves of the two prompt episodes of GRB 121217A acquired with BAT and GBM. The GBM triggered on the second peak which occurs at a time of $T_{0}+735\, s$, but has been shifted in this plot to coincide with the BAT $T_{0}$. Both light curves have been binned in time with a moving box of 5 seconds.}
  \label{fig:prompt_bat_and_gbm}
\end{figure}

The Burst Alert Telescope (BAT;~\cite{Barthelmy05a}) mounted on {\it Swift}~\cite{Gehrels04a} was triggered by GRB 121217 on 17th December 2012 at $T_{0}=$ 07:17:47 UT~\cite{Siegel12a}. {\it Swift} slewed immediately to the burst and the X-Ray Telescope(XRT;~\cite{Burrows05a}) began observing at $T_{0}+64.0\,\rm s$ until 15.6 days later~\cite{Evans12a}. The {\it Fermi} Gamma-ray Burst Monitor (GBM~\cite{Atwood09a}) was triggered by the second prompt emission of GRB 121217 on 17th December 2012 at 07:30:02 UT~\cite{Yu12a}. The prompt emission exhibits two main emission periods seperated by a quiescent period of $\sim700\,\rm s$. Both peaks have similar durations of $\Delta t_{\rm Peak1}=45\, \rm s$ and $\Delta t_{\rm Peak2}=33\,\rm s$, and fluences (10-1000 keV) of $f_{\rm Peak1}=4.8\pm0.6\times10^{-6}\,\rm erg\,cm^{-2}$ and $f_{\rm Peak2}=3.3\pm0.4\,\rm erg\,cm^{-2}$

\begin{figure}[htb]
\centering
\includegraphics[width=11cm]{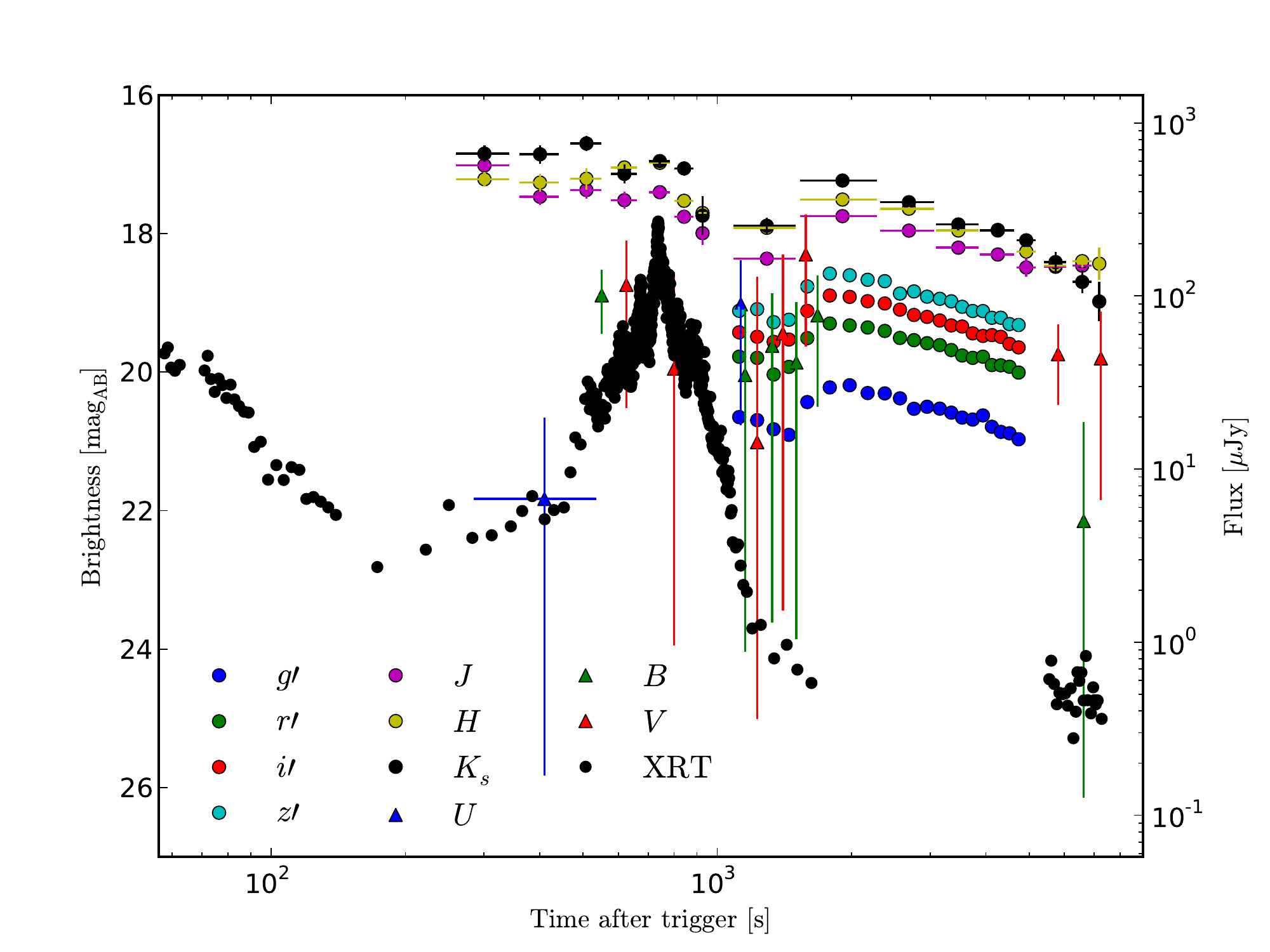}
\caption{The X-ray/optical-NIR light curve of GRB 121217A for the first day of observations, showing the $g'r'i'z'JHK$ filters of GROND, $UVB$ filters of UVOT and the XRT.}
\label{fig:lc}
\end{figure}

The Gamma-Ray burst Optical Near-infrared Detector (GROND~\cite{Greiner08a}) began observing the field of GRB 121217A at $T=T_{0}+210\,\rm s$ and located the optical counterpart~\cite{Elliott12b}. The follow up campaign lasted for $21\,\rm days$ until the afterglow was no longer detected, no underlying candidate host galaxy was discovered. The overall first day X-ray/optical wavelength light curve of GRB121217 can be seen in Fig.~\ref{fig:lc}. Additionally, the Ultra-Violet/Optical Telescope began observing 73 s after the trigger for 1.7 hours~\cite{Oates12a}.

\section{Results}

\subsection{Prompt Emission Broadband Spectrum}
\begin{figure}[htb]
\centering
\includegraphics[width=11cm]{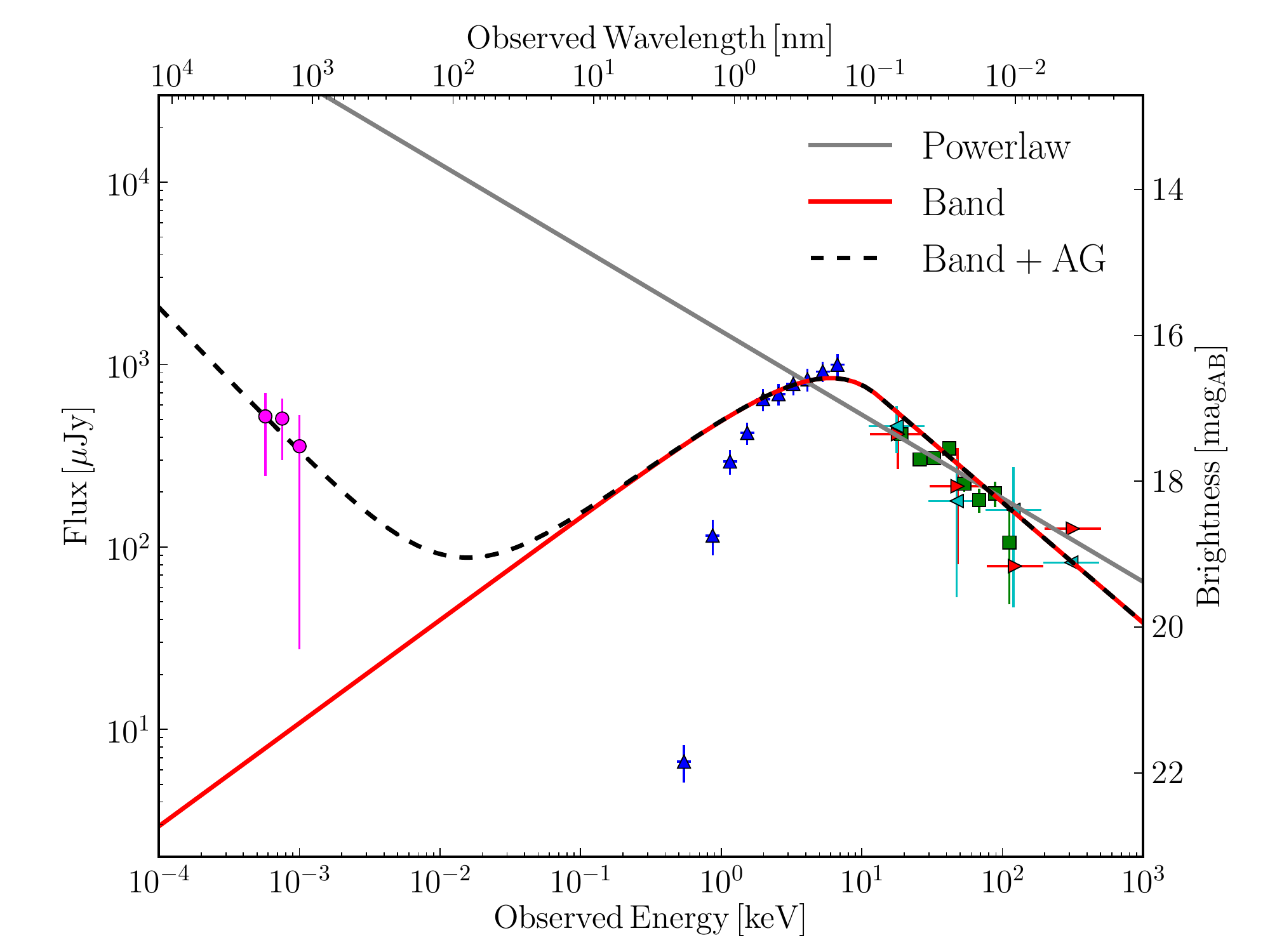}
\caption{Broadband SED at the second prompt emission, composed of BAT (green squares), XRT (blue upward-triangles), GBM Na and N9 detectors (red rightward-triangles and blue leftward-triangles, respectively), and the GROND $JHK$ channels (magenta).}
\label{fig:prompt_sed}
\end{figure}

We construct a broadband spectral energy (SED) distribution at the time of the second prompt emission, occuring at a mid-time of $T_{0}+735\pm10\,\rm s$, utilising the 3 GROND filters ($JHK$), {\it Swift}/BAT, {\it Swift}/XRT and {\it Fermi}/GBM (Fig.~\ref{fig:prompt_sed}). The SED is fit with a powerlaw and a Band model using the $\gamma$-ray emission, and the best-fit is extrapolated to the optical/NIR wavelengths. It is found that the powerlaw over-predicts the observed flux and the Band model under-predicts the flux. 

\subsection{Optical Rebrightening and Afterglow Component}
The optical light curve of GRB 121217, shows no significant rebrightening during the prompt emission of episode, changing by a factor of $\sim3.0$, which in comparison to such bursts as the naked eye burst, is negligible. Also, the X-ray emission increases by an order of $\sim100$ and peaks at the same time as the prompt emission. Such a scenario is viable if the afterglow component of the first prompt emission dominates, and/or the prompt spectrum is very hard, dampening the flux of the emission.

Therefore, we add an afterglow component to the observed prompt SED, with a standard slope of $\beta=0.8$ (assuming $F_(t,\nu)=t^{-\alpha}\nu^{-\beta}$), given that with only the NIR filters, an afterglow slope is difficult to constrain. This modifcation is seen to replicate well the observed SED and would suggest that the afterglow emission is dominant.

\subsection{Period of X-ray Rebrightening}

\begin{figure}[htb!]
\centering
\includegraphics[width=11cm]{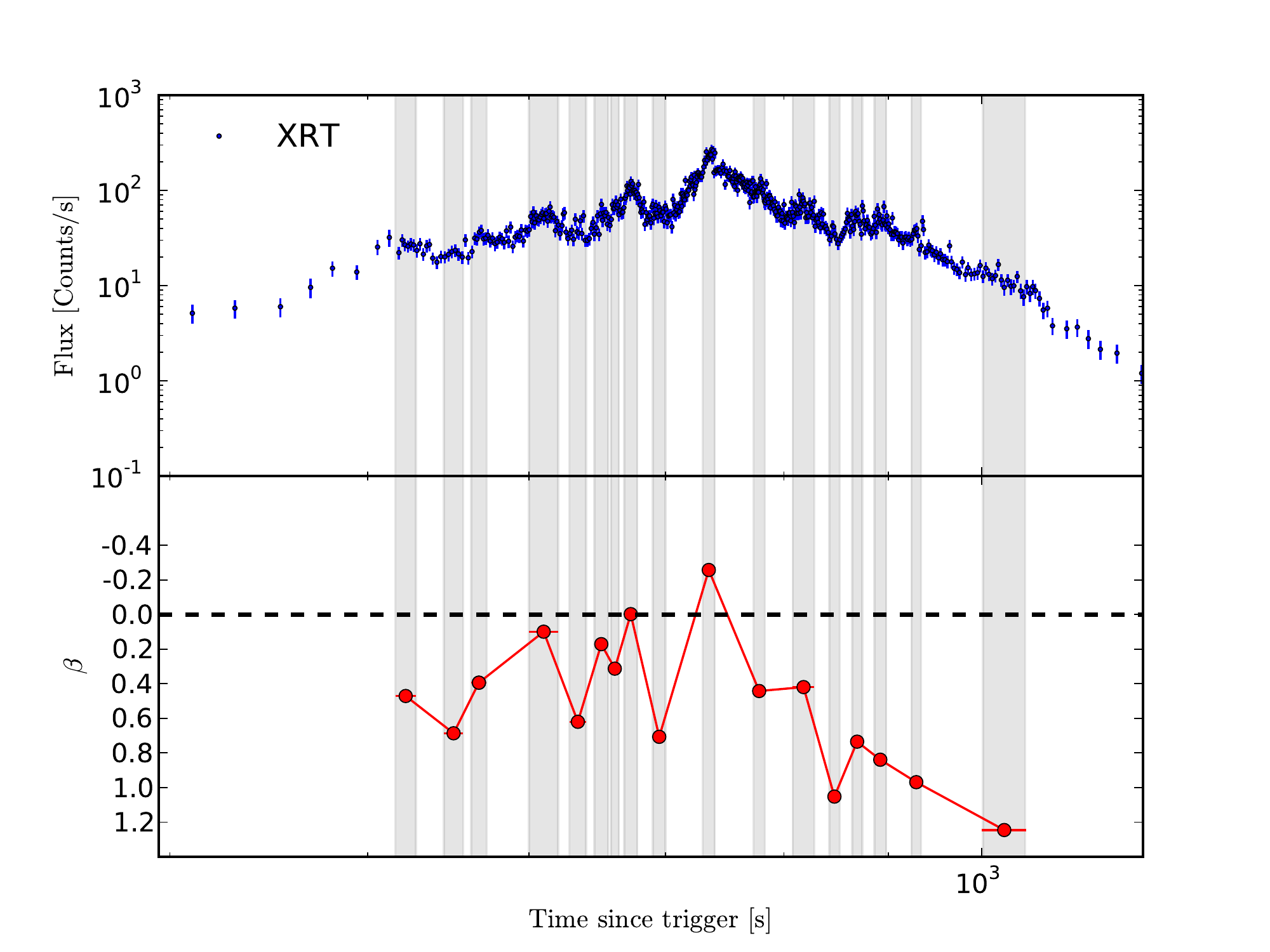}
\includegraphics[width=11cm]{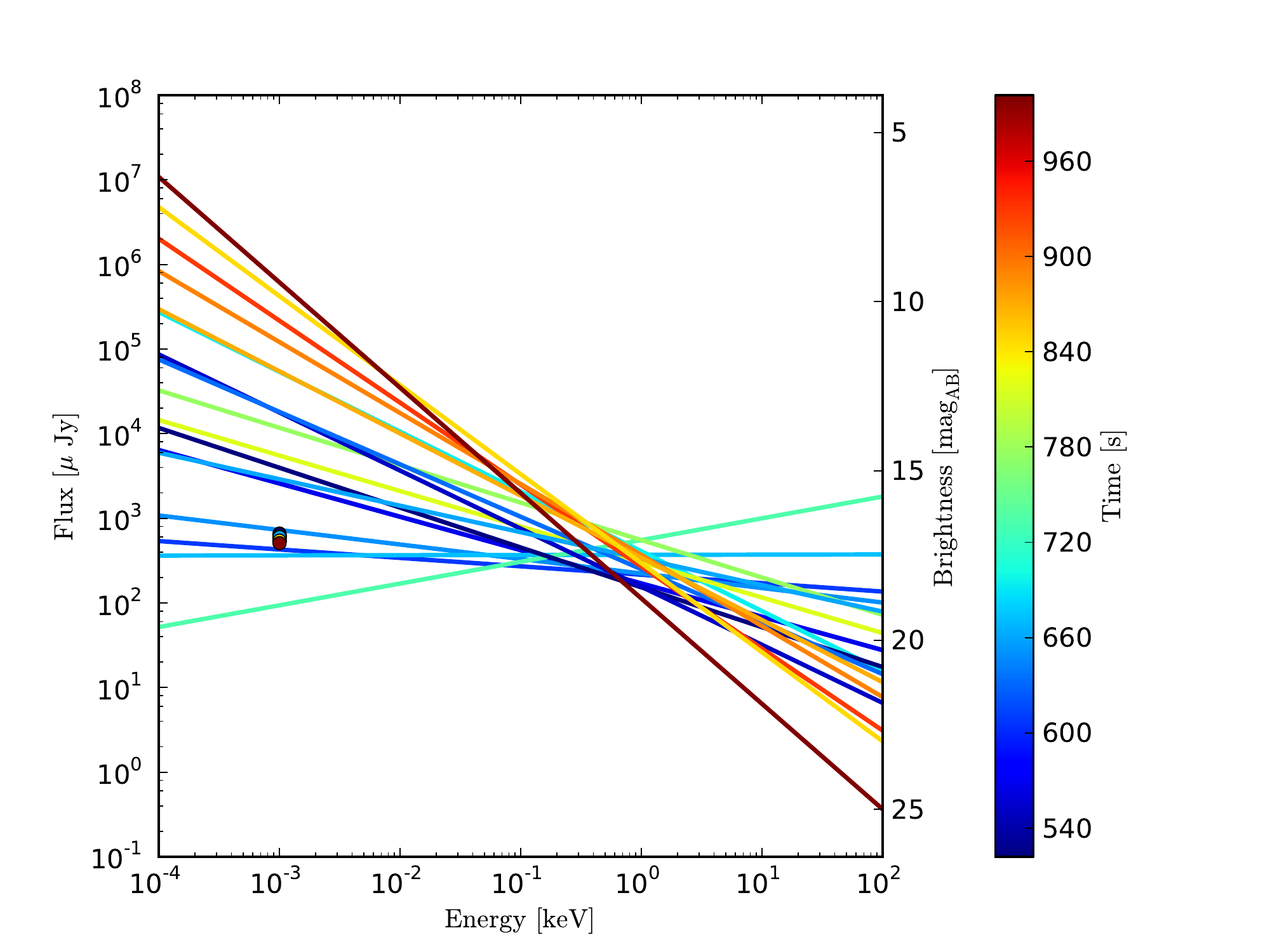}
\caption{{\bf Top}: The top panel shows a zoom of the X-ray light curve during the rebrightening episode. The lower panel shows the spectral slope of the X-ray spectrum obtained from the time slice depicted by the grey regions. {\bf Bottom}: The broadband spectra for each of the time slices obtained in the above figure, where the colour coding is the time of the slice. The dots depict the expected flux in the GROND $JHK$ bands for the same bands, using extrapolations from the best fit powerlaws obtained from the light curves.}
\label{fig:lc_pho}
\end{figure}

Before the second prompt emission, the X-ray light curve is seen to slowly rebrighten by a factor of $\sim100$, with several bump like features. We take several spectral slices of these bumps and ascertain the spectral slope of the X-ray spectrum, as seen in Fig.~\ref{fig:lc_pho}. Each of these bumps are spectrally soft ($\beta<0$), until the second prompt emission that is the hardest of all the slices ($\beta\sim-0.2$). In the framework of the internal shock model, one could think of several lower Lorentz factor shells tat are colliding but are not spectrally hard enough to be detected by any $\gamma$-ray detector, until eventually a fast enough shell collides. However, if soft enough, they would be observed in the optical wavelengths (Fig.~\ref{fig:lc_pho}), which is not the case. This again shows that the internal shock model cannot reproduce the observations, unless there exists a supression of the optical wavelength emission~\cite{Elliott13b}. For full details see~\cite{Elliott13c}.

\section{Conclusion}

We observed the {\it Swift/Fermi} burst GRB 121217 from two satellites and one telescope, with six different instruments covering the optical, X-ray and $\gamma$-ray wavelengths during a secondary prompt emission period. The optical emission exhibits no obvious rebrightening during this prompt episode, in contrast to cases such as the naked eye burst, and also in contrast to its own X-ray emission which increases by a factor of a hundred. 

Extrapolations of the internal shock model's best fit powerlaw overestimate the expected optical flux, and the extrapolation of the Band function under-predicts it, suggesting that neither model can reproduce the observations during the prompt emission. However, inclusion of an afterglow component as a result of the first prompt episode to the Band function can replicate the observations.

The number of observations of GRBs with simultaneous optical detections is beginning to become large enough to facilitate the compilation of hetrogeneous samples~\cite{Kopac13a}, however, there is still no consensus on the prompt emission mechanism. Further observations like that of GRB 121217A, especially in multiple filters, during the prompt emission of the GRB will help to further distinguish between the choices for the underyling mechanisms.


 

\end{document}